\documentclass[conference,screen]{IEEEtran}
\usepackage[misc]{ifsym}

\usepackage{booktabs}
\usepackage{multirow}
\usepackage{hyperref}
\usepackage{balance}

\usepackage{pifont}
\newcommand{\cmark}{\ding{51}}
\newcommand{\xmark}{\ding{55}}%

\usepackage{xcolor}
\usepackage{listings}

\definecolor{darkgreen}{rgb}{0, 0.44, 0.23}
\definecolor{lightgreen}{rgb}{0.25, 0.63, 0.4375}
\definecolor{darkblue}{rgb}{0.02, 0.16, 0.49}

\definecolor{redline}{rgb}{0.8,0.2,0.2}
\definecolor{greenline}{rgb}{0.02,0.5,0.02}

\definecolor{light-gray}{gray}{0.7}

\definecolor{cyanbox}{rgb}{0.28515625,0.75390625,0.71484375}

\definecolor{greenbox}{rgb}{0.5703125,0.8125,0.3125}

\definecolor{orangebox}{rgb}{1,0.6875,0.06640625}

\usepackage{listings, xcolor}

\definecolor{verylightgray}{rgb}{.97,.97,.97}

\lstdefinelanguage{Solidity}{
	keywords=[1]{anonymous, assembly, assert, balance, break, call, callcode, case, catch, class, constant, continue, constructor, contract, debugger, default, delegatecall, delete, do, else, emit, event, experimental, export, external, false, finally, for, function, gas, if, implements, import, in, indexed, instanceof, interface, internal, is, length, library, log0, log1, log2, log3, log4, memory, modifier, new, payable, pragma, private, protected, public, pure, push, require, return, returns, revert, selfdestruct, send, solidity, storage, struct, suicide, super, switch, then, this, throw, transfer, true, try, typeof, using, value, view, while, with, addmod, ecrecover, keccak256, mulmod, ripemd160, sha256, sha3}, 
	keywordstyle=[1]\color{blue}\bfseries,
	keywords=[2]{address, bool, byte, bytes, bytes1, bytes2, bytes3, bytes4, bytes5, bytes6, bytes7, bytes8, bytes9, bytes10, bytes11, bytes12, bytes13, bytes14, bytes15, bytes16, bytes17, bytes18, bytes19, bytes20, bytes21, bytes22, bytes23, bytes24, bytes25, bytes26, bytes27, bytes28, bytes29, bytes30, bytes31, bytes32, enum, int, int8, int16, int24, int32, int40, int48, int56, int64, int72, int80, int88, int96, int104, int112, int120, int128, int136, int144, int152, int160, int168, int176, int184, int192, int200, int208, int216, int224, int232, int240, int248, int256, mapping, string, uint, uint8, uint16, uint24, uint32, uint40, uint48, uint56, uint64, uint72, uint80, uint88, uint96, uint104, uint112, uint120, uint128, uint136, uint144, uint152, uint160, uint168, uint176, uint184, uint192, uint200, uint208, uint216, uint224, uint232, uint240, uint248, uint256, var, void, ether, finney, szabo, wei, days, hours, minutes, seconds, weeks, years},	
	keywordstyle=[2]\color{teal}\bfseries,
	keywords=[3]{block, blockhash, coinbase, difficulty, gaslimit, number, timestamp, msg, data, gas, sender, sig, value, now, tx, gasprice, origin},	
	keywordstyle=[3]\color{violet}\bfseries,
	identifierstyle=\color{black},
	sensitive=true,
	comment=[l]{//},
	morecomment=[s]{/*}{*/},
	commentstyle=\color{gray}\ttfamily,
	stringstyle=\color{red}\ttfamily,
	morestring=[b]',
	morestring=[b]"
}

\lstdefinestyle{mystyle}{
    basicstyle=\linespread{0.8}\footnotesize\ttfamily,
    breaklines=true,
    captionpos=b,
    numbers=left,
    numbersep=10pt,
    frame=tb,
    xleftmargin=\parindent,
    keywordstyle=\color{darkgreen},
    tabsize=2,
    commentstyle=\color{darkgreen},
}
\ifCLASSINFOpdf
  \usepackage[pdftex]{graphicx}
\else
\fi
\usepackage[cmex10]{amsmath}

\begin{document}
%
\title{Abusing the Ethereum Smart Contract Verification Services for Fun and Profit}



%
\author{\IEEEauthorblockN{Pengxiang Ma\IEEEauthorrefmark{1}\textsuperscript{\textsection},
Ningyu He\IEEEauthorrefmark{2}\textsuperscript{\textsection},
Yuhua Huang\IEEEauthorrefmark{1}, 
Haoyu Wang\IEEEauthorrefmark{1}\Letter, and
Xiapu Luo\IEEEauthorrefmark{3}}
\IEEEauthorblockA{\IEEEauthorrefmark{1}Huazhong University of Science and Technology}
\IEEEauthorblockA{\IEEEauthorrefmark{2}Peking University}
\IEEEauthorblockA{\IEEEauthorrefmark{3}The Hong Kong Polytechnic University}}

\maketitle
\begingroup\renewcommand\thefootnote{\textsection}
\footnotetext{The first two authors contribute equally.}
\endgroup

\begin{abstract}
Smart contracts play a vital role in the Ethereum ecosystem. Due to the prevalence of kinds of security issues in smart contracts, the smart contract verification is urgently needed, which is the process of matching a smart contract's source code to its on-chain bytecode for gaining mutual trust between smart contract developers and users. Although smart contract verification services are embedded in both popular Ethereum browsers (e.g., Etherscan and Blockscout) and official platforms (i.e., Sourcify), and gain great popularity in the ecosystem, their security and trustworthiness remain unclear. To fill the void, we present the first comprehensive security analysis of smart contract verification services in the wild. By diving into the detailed workflow of existing verifiers, we have summarized the key security properties that should be met, and observed eight types of vulnerabilities that can break the verification. Further, we propose a series of detection and exploitation methods to reveal the presence of vulnerabilities in the most popular services, and uncover 19 exploitable vulnerabilities in total. 
All the studied smart contract verification services can be abused to help spread malicious smart contracts, and we have already observed the presence of using this kind of tricks for scamming by attackers.
It is hence urgent for our community to take actions to detect and mitigate security issues related to smart contract verification, a key component of the Ethereum smart contract ecosystem.

\end{abstract}


%

\section{Introduction}
\label{sec:intro}
Ethereum, as one of the representative blockchain platforms, is regarded as a medal contender of Satoshi's Bitcoin. Its market cap peaked at \$540 billion in November 2021~\cite{ethmarketcap}. The success of Ethereum cannot omit the existence of tens of millions of deployed smart contracts on it. 

Specifically, smart contracts on Ethereum can be seen as scripts that will be executed once pre-defined conditions are met.
Alongside the characteristics of \textit{irreversibility} and \textit{determinacy} of blockchain, developers start to compose decentralized applications (DApps) through smart contracts, e.g., gambling games~\cite{game}, decentralized exchanges (DEXes)~\cite{uniswap}, and even decentralized autonomous organizations (DAOs) that can propose and consider proposals~\cite{dao}, where all participants are willing to and have to obey game rules hard-encoded into smart contracts.
Considering efficiency and I/O issues, Ethereum only stores a compact format, i.e., bytecode, of smart contracts within a decentralized database.

However, the unreadability of bytecode severely hampers the development of the ecosystem.
For example, on Ethereum, every single account is eligible to create and issue tokens that can be traded and exchanged. Because Ethereum does not require the uniqueness of all free-flowed tokens and provides a template for issuing tokens, it is hard to tell scam tokens from official or real ones by only auditing the bytecode. Xia et al.~\cite{xia2021trade} has identified over 10K scam token contracts, where scammers have gained a profit of at least \$16 million.
Such a gap between users' expectations and actual executed logic in unreadable bytecode urges the emergence of \textit{smart contract verification}, i.e., \textit{the process of matching a smart contract's source code to its on-chain bytecode}.

\begin{figure}[t]
    \centering
    \includegraphics[width=0.9\linewidth]{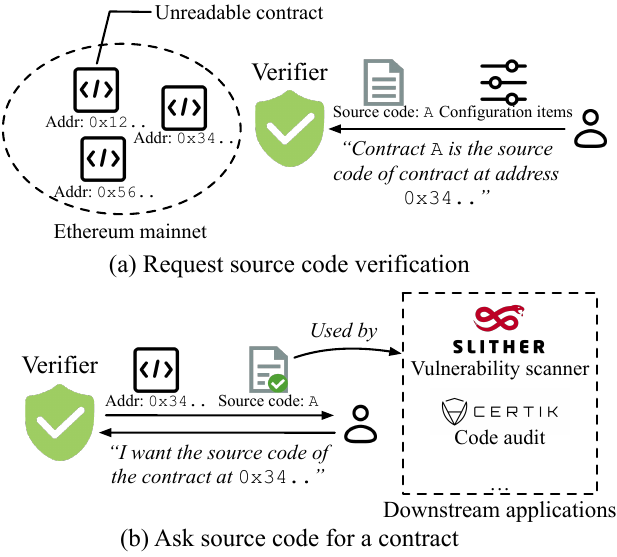}
    \vspace{-0.1in}
    \caption{Source code verification services in Ethereum ecosystem.}
    \vspace{-0.2in}
    \label{fig:ecosystem}
\end{figure}

Smart contract verification can be integrated into Ethereum browsers, e.g., Etherscan~\cite{etherscan} and Blockscout~\cite{blockscout}, or provided by the official platform, e.g., Sourcify~\cite{sourcify}.
As shown in Fig.~\ref{fig:ecosystem}, by providing the source code, a set of configuration items, and an address, the verification service can compile the given source code according to the configuration and compare it with the designated on-chain version (Fig.~\ref{fig:ecosystem} (a)).
To this end, all passed source code files are stored and displayed, as a piece of reference for users.
Once someone asks the source code of a contract, the verification service returns it if the source code is verified (Fig.~\ref{fig:ecosystem} (b)). The source code can be used in various kinds of downstream applications, e.g., vulnerability detection and code auditing. Indeed, many research studies~\cite{feist2019slither,grieco2020echidna} and industrial products take advantage of the smart contract verification services.
Intuitively, source code verification services will help gain mutual trust between smart contract developers and users, and boost a series of applications.

\textit{Are the smart contract verification services trustworthy?}
We should be aware that all our expectations are built upon the perfectly implemented source code verification services. However, no prior studies have considered whether these verification services work as expected.
Imagine such a situation where a malicious developer provides a seemingly harmless source code that can pass the smart contract verification of an on-chain bytecode, which however has been embedded with backdoors. 
Moreover, what if one's smart contract can be verified by arbitrary source code that is elaborately constructed by malicious competitors?

These inconsistencies can happen once there are vulnerabilities or bugs under any of the modules of the verification services, which can be abused by adversaries, leading to significant impact to the Ethereum ecosystem. 

\textbf{This Work.}
We take the first step to perform a comprehensive security analysis of Ethereum smart contract services in this paper. Based on the implementations of three mainstream smart contract verification services, i.e., Etherscan, Sourcify, and Blockscout, we first distill the general workflow of these verifiers and identify their key modules (see \S\ref{sec:source-code-verifier}).
Then we propose the key security properties that should be satisfied in these services, and summarize eight types of potential vulnerabilities that can break the verification once security properties are violated (see \S\ref{sec:security-analysis}).
Further, for each vulnerability, we propose a general detection method and the corresponding proof of concept (PoC) to illustrate how they can be exploited (see \S\ref{sec:detection}).
Our comprehensive analysis of the most popular smart contract verification services result in 19 vulnerabilities, and all the studied services can be attacked (see \S\ref{sec:evaluation}). After a timely and responsible disclosure to these verifiers, by the time of this submission, 15 vulnerabilities have been confirmed by official teams and 8 of them have been patched.
We further measure the impact of these vulnerabilities, and observe that tens of millions of deployed smart contracts can be abused by attackers, and hundreds of contracts have already been manipulated to perform fraud.

Our contributions can be summarized as follows:
\begin{itemize}
\item We depict the very first draw of the design and implementation of source code verifiers on Ethereum.
\item We uncover eight types of easily neglected but exploitable vulnerabilities hidden in verification services.
\item We propose the detection and exploitation methods against all these vulnerabilities, and uncovered 19 vulnerabilities in total. By the time of this writing, 15 vulnerabilities with high severity have been confirmed, and 8 of them have been patched.

\item We measure the impact of the vulnerabilities uncovered in this paper, and show that they can introduce great impact to the overall Ethereum ecosystem, i.e., tens of millions of contracts can be abused by attackers, and hundreds of contracts potentially have already been manipulated to perform scam activities.

\end{itemize}

\section{Background}
\label{sec:background}

\begin{figure}[t]
    \centering
    \includegraphics[width=0.7\linewidth]{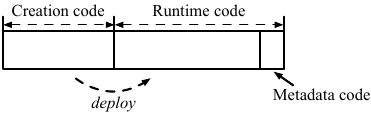}
    \vspace{-0.1in}
    \caption{The structure of a Ethereum smart contract.}
    \vspace{-0.2in}
    \label{fig:contract-structure}
\end{figure}

\subsection{Smart Contract}
\label{sec:background:contract}
Ethereum smart contracts are typically written in Solidity~\cite{solidityurl}, a high-level programming language specifically designed by Ethereum. Ethereum provides a specific compiler, named \textit{solc}, which takes source code written in Solidity as input, and generates a bytecode file, which can be executed by \textit{Ethereum Virtual Machine} (\textit{EVM}).
EVM bytecode corresponds to a series of opcodes, each of which is encoded as one byte, except for the \texttt{PUSH} opcode, which takes an immediate value as its operand~\cite{evmopcode}.
Further, EVM bytecode can be divided into three parts according to their functionalities, i.e., \textit{creation code}, \textit{runtime code}, and \textit{metadata code}, as shown in Fig.~\ref{fig:contract-structure}. Specifically, the creation code will be executed only once. It is responsible for deploying the corresponding runtime code on-chain, where the concrete executing logic of a smart contract, e.g., function implementations and variable declarations, is encoded.
Moreover, the fixed-length metadata code is part of the runtime code and not executable. It is the hash result of the metadata for reproducing the compilation~\cite{metadataintro}, e.g., ABI and solc version. Thus, the metadata code can be used as a key to retrieve and index the smart contract in a decentralized database. 
Note that, we often use the term, \textit{bytecode}, as a general one to refer all these three parts together.
When an initiator sends a piece of bytecode through a transaction, it will embed the bytecode into the \textit{input} field of the transaction with one or more parameters if the constructor requires.

Functions and variables in Ethereum smart contracts can be modified by several keywords~\cite{solidityurl}. For example, some visibility keywords, e.g., \texttt{private} and \texttt{external}, can be used to differentiate access control on functions and variables. Another keyword, named \texttt{immutable}, is introduced to modify hard-coded variables, which will be initialized till the contract is deployed by the creation code and cannot be updated afterwards.

\subsection{Smart Contract Verification}
\label{sec:background:source-code-verification}
\textit{Code Is Law}, is the core principle of Ethereum. 
Because on-chain smart contracts are non-updatable, behaviors of functions are determined and cannot be reverted. Moreover, due to the trustlessness among accounts (also smart contracts), the implementation of on-chain smart contracts captures everyone's attention. 
As we mentioned in \S\ref{sec:background:contract}, smart contracts are stored on-chain in the form of EVM bytecode. Due to its unreadability, it requires huge efforts to identify developers' original intention, or check whether unexpected behaviors (like attacks through backdoors and vulnerabilities) will occur.

Such an inconvenience urges the emergence of \textit{smart contract verification}\footnote{Note that we use \textit{smart contract verification} and \textit{source code verification} interchangeably in this paper.}, which is either a feature offered by Ethereum browsers (e.g., Etherscan~\cite{etherscan} and Blockscout~\cite{blockscout}) or a service provided by Ethereum official (e.g., Sourcify~\cite{sourcify}).
Specifically, anyone can upload a source code file and claim it is the original source code of an on-chain smart contract. The service will verify if the bytecode compiled from the given source code matches the one deployed on-chain. Once matched, users are able to query the source code files of verified contracts for further usage.

Note that, the matched verification results can be further divided into two types, i.e., \textit{partial match} and \textit{exact match}. To be specific, if the given source code can generate identical results with the on-chain one except for the metadata code, it is regarded as a \textit{partial match}. This is because some compilation options are inconsistent with the ones at the time of deployment. Otherwise, if all three parts can be matched exactly, it will be an \textit{exact match}.
However, this is only distinguished by Sourcify, i.e., an exact match can replace the existing partial match results~\cite{typeofmatch}. In contrast, Blockscout and Etherscan do not allow replacements of already user-verified contracts.
However, according to the statistics~\cite{ortnersmart}, only around 2\% of contracts on Etherscan have been verified, which leaves a huge attack surface for adversaries.

\section{Adversary Model}
\label{sec:adversary}

\begin{figure}[t]
    \centering
    \includegraphics[width=\linewidth]{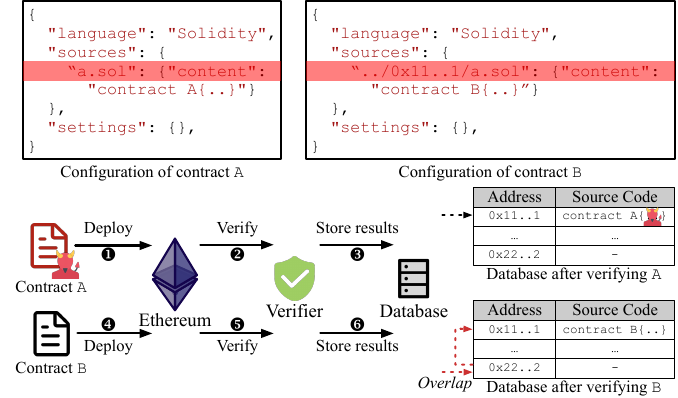}
    \vspace{-0.2in}
    \caption{A motivating example that exploits source code verification services.}
    \vspace{-0.2in}
    \label{fig:motivatingattack}
\end{figure}

\subsection{Motivating Example}
\label{sec:background:motivating}
If the smart contract verification services are exploited, the verified source code will not be inconsistent with the deployed on-chain bytecode. Fig.~\ref{fig:motivatingattack} illustrates a concrete motivating example that can achieve this goal.
As we can see, from \textbf{step 1} to \textbf{step 3}, a malicious user intentionally deploys an evil contract (\texttt{A}), requests a source code verification till the source code is stored in an address, \texttt{0x11...1} for example.
Then, to hide the backdoor, the attacker composes another confusingly similar contract but without the backdoor, named \texttt{B}.
From \textbf{step 4} to \textbf{step 6}, based on the contract \texttt{B}, the attacker repeats the above processes. However, at \textbf{step 5}, he provides a slightly modified configuration file. As the highlighted row indicates, he claims the path of \texttt{B} is \texttt{../0x11...1/a.sol}.
To this end, the source code of \texttt{B} overlaps the original one of \texttt{A} in the back-end database, which means users who visit \texttt{0x11...1} may obtain the source code of \texttt{B} with no explicit warnings. This vulnerability is uncovered in Sourcify by us (see \S\ref{sec:detection}).

\subsection{Adversary Model}
We assume that attackers can access the on-chain and off-chain data as normal users. Specifically, attackers can access all deployed smart contracts through a self-deployed node or Ethereum browsers, and request smart contract verification by providing necessary files.
Thus, according to whether an attacker requests smart contract verification on contracts deployed by himself or others, we can divide the consequences of exploiting smart contract verification services into two categories, i.e., \textit{competitive verification} and \textit{source code scam}.

\noindent
\textbf{Competitive Verification ($\mathcal{A}_1$).}
This consequence corresponds to verifying contracts deployed by other developers.
Given an arbitrary contract bytecode, malicious users can carefully craft a piece of source code that can successfully pass the verification process, by exploiting the vulnerabilities in the verifiers; even the original smart contract deployer does not intend to make his smart contract public. The verified source code can show malicious/ugly code that brings shame on the original projects. As for the \textit{competitive}, it can be explained in two ways.
On the one hand, a successful source code verification, even not requested by the deployer, has no chance to be replaced in Blockscout and Etherscan (see \S\ref{sec:background:source-code-verification}).
On the other hand, in Sourcify, if a smart contract is only partially verified (see \S\ref{sec:background:source-code-verification}), attackers can construct an exact match to competitively replace the original verifying results.

\noindent
\textbf{Source Scam ($\mathcal{A}_2$).}
Under this scenario, the attacker is the developer of the contract (see our motivating example).
Attackers can exploit the vulnerabilities in verifiers to hide their original intentions. For example, they can deploy contracts with backdoors, but forge a piece of source code without backdoors that can bypass the verification. 
Consequently, the disguised code gains users' trust, which is however a trap.

Except for \textit{direct} influence on Ethereum users (like $\mathcal{A}_1$ and $\mathcal{A}_2$), such a behavior can also lead to serious consequences.
For example, lots of downstream applications, e.g., source-code level vulnerability detectors (slither~\cite{feist2019slither} and echidna~\cite{grieco2020echidna}) and smart contract auditing companies, rely on verified source code retrieved from APIs of these three platforms.
Therefore, if the verified source code is contaminated, the reliability of services provided by all downstream applications would be affected.

\section{Source Code Verifier}
\label{sec:source-code-verifier}

We first investigate the general workflow of three mainstream source code verifiers, i.e., Etherscan, Sourcify, and Blockscout. Specifically, Sourcify~\cite{sourcify} and Blockscout~\cite{blockscout} are open-source projects, thus we conduct a comprehensive code audit on their documentation and implementations. As for Etherscan, whose implementation is proprietary, we compose a set of contracts to perform black-box testing.

\begin{figure*}[t]
    \centering
    \includegraphics[width=0.9\linewidth]{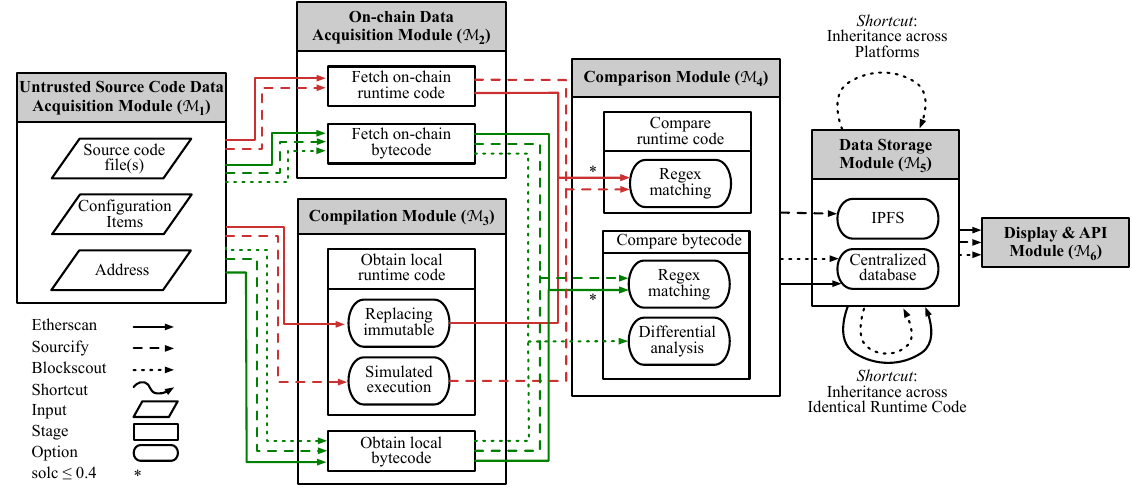}
    \vspace{-0.1in}
    \caption{Architecture and workflow of the source code verifier, where \textcolor{redline}{red} and \textcolor{greenline}{green} lines refer to runtime code and bytecode related processes, respectively.}
    \vspace{-0.2in}
    \label{fig:module}
\end{figure*}

\subsection{Overview}
\label{sec:source-code-verifier:overview}

Fig.~\ref{fig:module} illustrates the workflow and architecture of the source code verifier.
Specifically, it takes source code files, configuration items, and an address as inputs. According to the input configuration items, the verifier can compile the given source code in a deterministic way.
Then, the verifier fetches both the bytecode and the runtime code according to the input address, and compares if they are identical to the just compiled ones. 
Once matched, the provided source code would be labeled as \textit{verified} and demonstrated to users who query the source code of the address via the APIs exposed by verifiers. Otherwise, it means that either the given source code is not the implementation of the on-chain bytecode, or some configuration items are inconsistent.

\subsection{Components of Verifier}
\label{sec:source-code-verifier:components}
Diving into the implementation of smart contract verifiers, there are generally six major components, denoted from $\mathcal{M}_1$ to $\mathcal{M}_6$, which will be depicted in the following.

\subsubsection{Untrusted Source Code Data Acquisition Module ($\mathcal{M}_1$)}
\label{sec:source-code-verifier:components:m1}
$\mathcal{M}_1$ is responsible for obtaining those untrusted data, i.e., source code file(s), configuration items, and an address, from the verifying requester.
As for providing multiple source code files, it is because a \textit{decentralized application} (\textit{DApp}) in Ethereum may require multiple files to achieve complex functionality.
To enable compiling a DApp with hierarchical structure among source code files, solc additionally allows users to compact all source code files and their paths into a single file with JSON format.
Additionally, once a contract is compiled by solc, a set of configuration items are generated automatically. They indicate the metadata during the compilation, e.g., function signatures and optimizer options. Finally, the one who requests verification should also provide an address, the bytecode stored in which is regarded as the verification target. Some extra information may also be required, depending on specific service providers.
For example, Blockscout requires the contract name because multiple contracts can exist within a single file, while only one of them is the entry.

\subsubsection{On-chain Data Acquisition Module ($\mathcal{M}_2$)}
\label{sec:source-code-verifier:components:m2}
This module retrieves necessary on-chain data, i.e., the runtime code, and the bytecode that is located in the very first transaction of the contract, by querying RPCs exposed by client nodes or APIs of Ethereum databases. Note that, different verifiers have diverse verification strategies, thus the on-chain data fetch is conducted on necessary.
Specifically, Etherscan requires both bytecode and runtime code should be identical to the on-chain ones, while Sourcify only requires either one of them. In addition, Blockscout only conducts verification on bytecode. To obtain on-chain runtime code, $\mathcal{M}_2$ queries the standard RPC, named \texttt{eth.getCode()}, which is exposed by client nodes.
As for obtaining bytecode, it requires more effort. $\mathcal{M}_2$ firstly queries some third-party databases, e.g., Etherscan~\cite{etherscandatabase}, to obtain the hash of the creator transaction to the address, and extract the bytecode from the \texttt{input} field of the transaction by an RPC, named \texttt{eth\_getTransactionByHash}.

\subsubsection{Compilation Module ($\mathcal{M}_3$)}
\label{sec:source-code-verifier:components:m3}
$\mathcal{M}_3$ invokes solc to compile and obtain both bytecode and runtime code for the following comparison processes.
According to source code files and configuration items obtained by $\mathcal{M}_1$, solc can compile in a deterministic way. Even so, the directly obtained runtime code may be unusable for the following verification. The reason lies in that there may be \textit{immutable} variables declared (see \S\ref{sec:background:contract}), whose values are dynamically determined by executing the creation code.
After a thorough investigation, we find two ways are adopted by verifiers to resolve this issue, i.e., \textit{directly replacing immutable variables}, and \textit{simulately executing creation code}~\cite{sourcifysimulation}.

As for the former method, during compilation, it leaves all immutable variables as blanks and records all their offsets. Then, it fetches the corresponding bytes from the on-chain runtime code according to recorded offsets, and fills them in the blanks of the local runtime code. Hence, the compiled immutable values are definitely identical to the on-chain ones, which \textit{does not achieve the purpose of verification}.
As for the latter method, to obtain the corresponding runtime code with immutable variables declared, $\mathcal{M}_3$ invokes \texttt{eth\_call} exposed by client node to simulate the behavior of creation code.
This method plays well under normal circumstances, because \texttt{eth\_call} interprets each opcode in the creation code in order.
For example, \texttt{int immutable public a = 20**23}, where \texttt{a} is an immutable variable, whose value is calculated and determined during the deployment.

\subsubsection{Comparison Module ($\mathcal{M}_4$)}
\label{sec:source-code-verifier:components:m4}
$\mathcal{M}_4$ is responsible for comparing on-chain fetched data and the locally compiled one, in terms of both bytecode and runtime code.
As we stated in $\mathcal{M}_2$, verifiers adopt different strategies, i.e., Blockscout only needs bytecode, while Etherscan requires both bytecode and runtime code matching. Interestingly, Sourcify only requires either of them during the comparison.
Note that, on-chain bytecode is obtained from the \texttt{input} field of the creation transaction, whose bytecode may be followed by parameters if the constructor requires (see \S\ref{sec:background:contract}). Thus, if verifiers do not require requesters to provide initial values of these parameters, a commonly adopted strategy is to determine whether the locally compiled bytecode is a prefix of the creation transaction input.

Except for that, comparing bytecode or runtime code is still challenging due to the existence of metadata.
As Fig.~\ref{fig:contract-structure} illustrates, the metadata often locates at the end of the runtime code. Moreover, though we can offer a set of configuration items to generate identical metadata, it is still difficult to guarantee an identical set of configuration items with the ones adopted when actually deploying on-chain bytecode. Thus, removing the metadata code before performing contract comparison is the best choice to avoid verification failure.

These three platforms adopt distinct methods to identify and remove metadata code.
Specifically, because the metadata code is wrapped by a fixed length bytes~\cite{solidityurl}, Sourcify directly identifies the pattern through regex matching at the tail of the runtime code.
However, some contracts may have multiple metadata in the middle of runtime code, like factory contracts that can deploy other contracts by themselves, where only removing the tailing metadata code is insufficient.
Thus, Blockscout adopts another method. It chooses and updates some configuration items, which only changes the metadata code while keeping runtime logic intact. Then, it identifies all metadata by observing which bytes are updated due to such a deliberate modification. As for the proprietary Etherscan, we intentionally send some factory contracts to observe if it can identify multiple metadata code. The results show that the verification cannot pass due to unidentified metadata when solc \texttt{$\leq$ 0.4}, but plays opposite when solc is greater than \texttt{0.4}. Thus, at least we can conclude that when solc \texttt{$\leq$ 0.4}, Etherscan adopts the same method as Sourcify.

\subsubsection{Data Storage Module ($\mathcal{M}_5$)}
\label{sec:source-code-verifier:components:m5}
After completing the above comparison, all uploaded files are stored for the subsequent querying and display. Thus, $\mathcal{M}_5$ is accountable for storing them in a permanent database.
For Etherscan and Blockscout, they store the source code files in their owned and centralized back-end server. However, users may be concerned about such a centralization issue in \textit{decentralized} blockchain platforms. Thus, Sourcify adopts IPFS~\cite{ipfs}, a decentralized file system that can be accessed by anyone who maintains an IPFS node, to enhance its confidence for users.

\subsubsection{Display \& API Module ($\mathcal{M}_6$)}
\label{sec:source-code-verifier:components:m6}
Through $\mathcal{M}_6$, users can access all necessary information of the corresponding source code files when asking for a contract of an address.
The displayed information typically includes EVM version, solc version, contract name, and linked library addresses.
To make it more clear, Sourcify directly shows a metadata file to users which contains all configuration items.

\subsection{Shortcuts}
\label{sec:source-code-verifier:shortcuts}
All verifiers implemented by these three platforms generally follow the design and workflow we mentioned above. However, we found some shortcuts, i.e., skipping some modules when a condition meets, exist in their implementations. The purpose of these shortcuts is to alleviate the workload or save computing resources of verifiers.
We detail the found two shortcuts in the following.

\noindent
\textbf{Source Code Inheritance across Identical Runtime Code.}
Intuitively, if two on-chain runtime code are identical, we can assume there is nothing different in their implementations of the non-constructor part. 
Therefore, in Etherscan and Blockscout, if a contract is verified by a set of source code files, all other on-chain runtime code that are identical to this one are also linked to these files. Such a shortcut provides an automatic verification for \textit{factory contracts}. Take a famous decentralized finance application, Uniswap~\cite{uniswap}, as an example, which allows arbitrary token exchanges at an interest rate calculated by supply and demand.
Users could create an exchangeable token pair by invoking \texttt{create\_pair} in its factory contract. Because all created contracts are identical, the previously verified source code files can be directly inherited to newly created ones without any verification request.

\noindent
\textbf{Source Code Inheritance across Platforms.}

To avoid resource consumption, Blockscout recognizes the verification results of Sourcify, which does not need further manual verifying requests on the former platform. In other words, against an address, users can only request verification on Sourcify, Blockscout automatically inherits the source code files uploaded to the address. Note that, there is no automatic results relay from or to Etherscan, which still requires manual verification requests from users for each contract.

\section{Security Properties and Vulnerabilities}
\label{sec:security-analysis}

We next present the key security properties that should be satisfied in \S\ref{sec:security-analysis:properties}.
Further, we discuss the vulnerabilities that would appear if the security properties are violated in \S\ref{sec:security-analysis:violate}.

\subsection{Security Properties}
\label{sec:security-analysis:properties}
As we mentioned in \S\ref{sec:adversary}, an attacker can access all on-chain data, e.g., contract bytecode and transaction history, as normal Ethereum users. Moreover, source code verification services do not require authentication before requesting verification or querying verification results.
Thus, a set of security properties ($\mathcal{P}$) should be guaranteed to protect the normal functionality of source code verification services.
As shown in Table~\ref{table:properties}, we propose 3 security properties, by carefully considering options in modules and workflow of verifiers.

\begin{table}[t]
\centering
\caption{Security Properties for Source Code Verifiers.}

\label{table:properties}
\begin{tabular}{@{}ccc@{}}
\toprule
\textbf{}       & \textbf{Security Property} & \textbf{Related Modules} \\ \midrule
$\mathcal{P}_1$ & Data Integrity             & $\mathcal{M}_3$, $\mathcal{M}_4$                         \\
$\mathcal{P}_2$ & Data Consistency           & $\mathcal{M}_2$, $\mathcal{M}_3$, $\mathcal{M}_4$, $\mathcal{M}_5$                         \\
$\mathcal{P}_3$ & Unambiguous Reference      & $\mathcal{M}_6$          \\ \bottomrule
\end{tabular}
\vspace{-0.2in}
\end{table}

\noindent
\textbf{Data Integrity ($\mathcal{P}_1$).}
\textit{Data integrity} refers to that data cannot be tampered with. Under the scenario of source code verification services, it refers to that malicious users cannot arbitrarily forge a piece of source code to bypass the verification through reverse engineering on-chain bytecode or utilizing features of verifiers.
To meet this property, verifiers should be robust to the following threats.
First, \textit{threats to the compilation module ($\mathcal{M}_3$)}. Features in the compilation module may allow attackers to generate source code files according to on-chain bytecode. For example, both the verbatim function~\cite{verbatimsol} and loose assembly~\cite{loosesol} in solc can assist developers to output a sequence of designated opcodes, which can be maliciously used by attackers.
Second, \textit{threats to the comparison module ($\mathcal{M}_4$)}. During the comparison stage, as we mentioned in \S\ref{sec:source-code-verifier:components:m4}, some options are enabled to enhance the user-friendliness, like only considering the prefix of local compiled bytecode. Such options can be utilized by attackers to hamper the data integrity.
If data integrity breaks, attackers can forge source code of on-chain bytecode, which results in the $\mathcal{A}_1$ consequence.

\noindent
\textbf{Data Consistency ($\mathcal{P}_2$).}
\textit{Data consistency} refers to a property that the semantics between verified source code files and on-chain bytecode should be consistent. Three types of threats can break this property.
First, \textit{part of source code is not considered}. For example, in $\mathcal{M}_3$, the implementation of invoked functions in linked libraries is not included in the caller. Thus, the lack of a recursive verification on such functions can be utilized by attackers to hide backdoors.
Second, \textit{part of runtime logic is neglected}. Before the comparison in $\mathcal{M}_4$, all metadata code should be identified and removed. However, if a piece of runtime logic is mistakenly marked as metadata, its semantics is overlooked during comparison.
Third, \textit{stored data is replaced}. After a successful verification, the uploaded source code files are stored in $\mathcal{M}_5$. The stored source code may be overlapped unexpectedly or intentionally by tampering file systems. For example, the \texttt{create2} opcode~\cite{create2risk} can be maliciously utilized to replace the on-chain bytecode on a designated address.
Consequently, such an inconsistency between the on-chain bytecode and the uploaded source code files can result in the $\mathcal{A}_2$ consequence.

\noindent
\textbf{Unambiguous Reference ($\mathcal{P}_3$).}
\textit{Unambiguous reference} is mainly associated with $\mathcal{M}_6$. Specifically, the unambiguity means that all displayed source code files should not be vague, and data returned by APIs should be complete and explicit.
Because lots of downstream applications solely depend on data exposed by $\mathcal{M}_6$, like the example shown in Fig.~\ref{fig:ecosystem}. Only unambiguous data can guarantee determined results during handling the corresponding source code. If the effectiveness of downstream applications is influenced, or users who query source code of an address are misled, it may lead to the consequence $\mathcal{A}_2$, i.e., source code scam.

\begin{figure}[t]
    \centering
    \includegraphics[width=\linewidth]{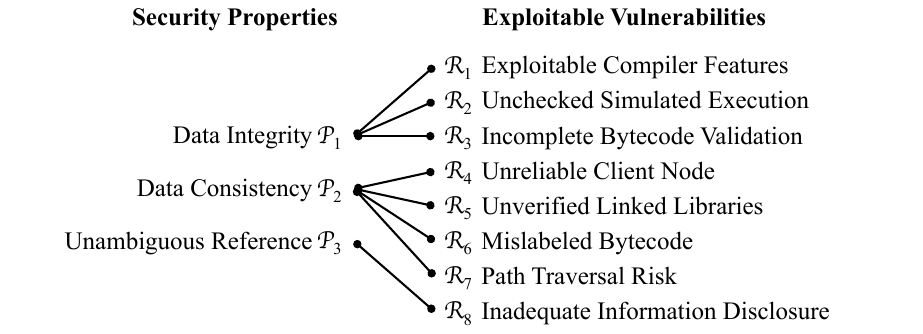}
    \vspace{-0.3in}
    \caption{Relationship between security properties ($\mathcal{P}$) and exploitable vulnerabilities ($\mathcal{R}$) in source code verifiers.}
    \vspace{-0.2in}
    \label{fig:properties-vulnerabilities}
\end{figure}

\subsection{Vulnerabilities that Violate Security Properties}
\label{sec:security-analysis:violate}
If security properties are violated, exploitable vulnerabilities shall emerge. We identified eight kinds of vulnerabilities that could lead to the violation of security properties, by resorting to the \texttt{solc} documentation and the implementation of the verifiers. Fig.~\ref{fig:properties-vulnerabilities} shows the relationship between security properties and exploitable vulnerabilities we uncovered. Note that the detection and exploitation of these vulnerabilities will be presented in \S\ref{sec:detection}.

\noindent \textbf{Overview}
As shown in Fig.~\ref{fig:properties-vulnerabilities}, we have summarized eight kinds of exploitable vulnerabilities in total.
Specifically, $\mathcal{P}_1$ corresponds to three exploitable vulnerabilities, which break the data integrity, i.e., attackers can forge a piece of source code according to unreadable on-chain bytecode to bypass the verification mechanism.
Moreover, $\mathcal{P}_2$ can lead to four vulnerabilities to damage the data consistency. In other words, by exploiting these vulnerabilities, attackers can somehow make the on-chain deployed bytecode inconsistent with the uploaded source code files.
At last, $\mathcal{P}_3$ is only related to one vulnerability, i.e., $\mathcal{R}_8$. Attackers can conduct source code scam through the ambiguous reference exposed by APIs from $\mathcal{M}_6$.

\begin{lstlisting}[language=Solidity, caption={The verbatim feature in YUL.}, label={lst:r1}]
object "verbatim" {
  code {
    verbatim_0i_0o(hex"60806040")
  }
}
// actually deployed bytecode: 60806040fe
\end{lstlisting}
\vspace{-0.1in}

\subsubsection{Exploitable Compiler Features ($\mathcal{R}_1$)}
\label{sec:risk:r1}
In $\mathcal{M}_3$, features of solc give attackers opportunities to generate user-controlled bytecodes that violate $\mathcal{P}_1$.
Specifically, such features include \textit{YUL}~\cite{yul} (an intermediate language) and \textit{loose inline assembly}~\cite{loosesol}, both of which can be inlined in source code.
For example, Listing~\ref{lst:r1} illustrates a code snippet written in YUL. As illustrated at L3, users can directly designate a list of opcodes via the built-in function \texttt{verbatim\_0i\_0o}.
Though according to the implementation of solc, an invalid opcode \texttt{fe} is appended to the final deployed contract (L6), it still reflects how users can flexibly control the deployed bytecode via features of the compiler.
Likewise, when solc \texttt{$\leq$ 0.4}, users can adopt loose inline assembly to generate arbitrary opcode sequences.

\subsubsection{Unchecked Simulated Execution ($\mathcal{R}_2$)}
\label{sec:risk:r2}
Simulated execution is an option in $\mathcal{M}_3$, which takes whatever it passes as a creation code, executes it locally, and regards the returned bytecode as the corresponding runtime code.
Such a process may violate $\mathcal{P}_1$ if attackers utilize this feature to construct the return value.
As we mentioned in \S\ref{sec:source-code-verifier:components:m3}, simulated execution can be used to assign immutable variables. However, such a simulation can be regarded as an \texttt{eval()} function, which is proven notoriously unsafe in multiple languages~\cite{richards2011eval}.
Intuitively, it can be utilized by an elaborately constructed creation code.
For example, in normal creation code, the pointer of the runtime code is acquired by the last \texttt{codecopy} opcode. Then, the pointed data, i.e., runtime code, is returned to EVM.
However, Solidity does not check if the pointer points to the correct runtime code at all, which means that attackers can construct a creation code within which returns a wrong pointer.
Thus, they can assemble a piece of bytecode, e.g., the victim's bytecode, pointed by a pointer, and deliberately return it before the actual \texttt{codecopy} in creation code.
Consequently, the victim's bytecode is regarded as the runtime code of this constructed source code.

\begin{lstlisting}[language=Solidity, caption={A contract that requires a constructor parameter.}, label={lst:r6}]
contract Sample{
  uint public var;
  constructor (uint a){
    var = a;
  }
}
// Local ByteCode:
// |--bytecode--|
// 60806040..0033
// Creation Transaction Input:
// |--bytecode--|--Consturctor Paramert--|
// 60806040..00330000..0000000000000000001
\end{lstlisting}
\vspace{-0.1in}

\subsubsection{Incomplete Bytecode Validation ($\mathcal{R}_3$)}
\label{sec:risk:r3}
Considering user-friendliness, as we introduced in \S\ref{sec:source-code-verifier:components:m4}, verifiers only perform a prefix check when comparing the bytecode in $\mathcal{M}_4$. Such a strategy can be utilized by attackers if the prefix identification goes wrong, which violates $\mathcal{P}_1$. Specifically, constructors may take arguments, like the code snippet shown in Listing~\ref{lst:r6}. The constructor at L3 takes an argument \texttt{a}, whose initial value should be appended to the \texttt{input} field of the creation transaction. The existence of arguments makes a direct comparison always mismatched, i.e., directly comparing L9 and L12.
Thus, verifiers adopt \textit{prefix matching}, i.e., if the locally compiled bytecode (L9) is the prefix of the on-chain one (L12). If it is, they consider the local one as verified.
However, this strategy can be exploited if there lacks a complicated examination.
For example, a prefix of L9 can also be the prefix of L12 to pass such a comparison process. Furthermore, if verifiers do not take an empty string into consideration, the prefix check may also be bypassed.
Consequently, attackers can construct a source code to generate a prefix of the victim's bytecode to bypass the verification.

\subsubsection{Unreliable Client Node ($\mathcal{R}_4$)}
\label{sec:risk:r4}
The status and enabled features of the node that verifiers depend on may violate $\mathcal{P}_2$.
As we mentioned in \S\ref{sec:source-code-verifier:components:m2}, $\mathcal{M}_2$ is responsible for fetching on-chain data from Ethereum. However, if it relies on a single client node that happens to be stuck into troubles, data consistency cannot be guaranteed. 
For instance, client nodes may suffer the chain reorganization issue~\cite{neuder2021low}. This leads to returning unconfirmed on-chain bytecode, which may impact the following on-chain verification. Additionally, the EVM version adopted by the verifier-dependent node may also lead to risks.
Specifically, in \texttt{V0.6.2}~\cite{ethcreate2}, Ethereum has introduced a new opcode, named \texttt{create2}. It was originally designed to deploy a smart contract on a predetermined address for developers. However, this opcode could be abused to re-deploy a piece of runtime code on an existing address, which may invalidate the original source code verification results~\cite{create2risk}. 

\begin{lstlisting}[language=Solidity, caption={The difference between linked and embedded libraries.}, label={lst:r3}]
library A_embedded{ // embedded library
  function foo() internal returns(uint){
    ...
}
}
library A_linked{ // linked library
  function bar() public returns(uint){
    ...
}}
\end{lstlisting}
\vspace{-0.1in}

\subsubsection{Unverified Linked Libraries ($\mathcal{R}_5$)}
\label{sec:risk:r5}
The implementation of invoked linked libraries is not considered by verifiers, which intuitively violates $\mathcal{P}_2$. Specifically, in order to support modular design when developing DApps, developers can write part of the contract logic into a \textit{library contract}.
Two kinds of libraries exist, i.e., \textit{linked library} and \textit{embedded library}, whose distinctions are subtle~\cite{libraryintro}.
For example, Listing~\ref{lst:r3} defines an embedded library (\texttt{A\_embedded}) and a linked library  (\texttt{A\_linked}), respectively. The only difference lies at L2 and L6, i.e., the function is modified by \texttt{public} or \texttt{internal}.
Except for declaration, these two kinds of libraries are encoded in different ways.
Specifically, a call to functions in linked libraries is encoded as an opcode \texttt{delegatecall}, which takes the invoked library's address and the function signature of the callee as arguments. This means that the logic of the callee does not appear in the caller's bytecode.
Conversely, as for embedded libraries, the bytecode of callee is directly included into the caller.
Such a slight difference may mislead users that all invoked libraries are also verified. This could result in exploitable loopholes when invoking linked libraries, whose implementations are not recursively verified.

\subsubsection{Mislabelled Bytecode ($\mathcal{R}_6$)}
\label{sec:risk:r6}
Before performing comparison in $\mathcal{M}_4$, verifiers ignore certain types of bytecode for user-friendliness and to avoid unnecessary verification failure. However, intentionally neglecting bytecode can lead to inconsistency issues, i.e., violating $\mathcal{P}_2$.
Two types of bytecode lie in this scope, i.e., \textit{metadata} and \textit{linked library placeholders}.

The necessity of removing all metadata code before $\mathcal{M}_4$ is detailed in \S\ref{sec:source-code-verifier:components:m4}. However, a problematic metadata code extraction may leave metadata behind or extract innocent bytecode as metadata, leading to a failed verification or verification bypass, respectively.
Specifically, the current pattern of metadata code is defined as~\cite{solidityurl}:
\begin{align*}
&\texttt{0xa2}\\
&\texttt{0x64}\ \texttt{i}\ \texttt{p}\ \texttt{f}\ \texttt{s}\ \texttt{0x58}\ \texttt{0x22}\ <\text{34\ bytes\ IPFS\ hash}>\\
&\texttt{0x64}\ \texttt{s}\ \texttt{o}\ \texttt{l}\ \texttt{c}\ \texttt{0x43}\ <\text{3\ byte\ solc\ version}>\\
&\texttt{0x00}\ \texttt{0x33}
\end{align*}
, where all letters are encoded by ASCII in bytes. To extract all metadata code, as we mentioned in $\mathcal{M}_4$, two ways are adopted by current verifiers, i.e., \textit{regex matching} and \textit{differential extracting}.
Specifically, no matter where a metadata code locates, regex matching is supposed to be effective and efficient.
Considering the possibility of updating the metadata pattern, Blockscout adopts differential extracting to filter all metadata out by inserting meaningless configuration items (see \S\ref{sec:source-code-verifier:components:m4}). However, if these configuration items are used by attackers to cause the runtime logic to change as well, the data inconsistency ($\mathcal{P}_2$) is violated.

Linked library placeholders should also be removed before $\mathcal{M}_4$.
To be specific, during the compilation, solc firstly replaces each invoked linked library address with a 40 bytes long placeholder. If users require verification on Etherscan or Blockscout, both verifiers ask users to provide concrete invoked addresses to replace placeholders.
Unlike them, Sourcify provides a more user-friendly solution. Sourcify identifies placeholders, takes each of them as a regex pattern, and conducts a regex match to replace them according to the on-chain bytecode at the same offsets.
However, if the placeholder can be constructed by attackers, runtime logic may be mislabelled, which violates the data consistency as well.

\subsubsection{Path Traversal Risk ($\mathcal{R}_7$)}
\label{sec:risk:r7}

Another noteworthy but easily overlooked risk needs to be mentioned, namely \textit{path traversal risk}, which could result in a violation of $\mathcal{P}_2$.
Specifically, to enable compiling a complicated DApp that consists of multiple source code files, solc allows users to compact source code files and their corresponding paths in a file with JSON format. This allows malicious users to designate the path of each file at will.
To this end, if the designated path is not sufficiently validated, it is very likely that an attacker can utilize this feature to construct an arbitrary path traversal, affecting the security of the source code verifiers.

\subsubsection{Inadequate Information Disclosure ($\mathcal{R}_8$)}
\label{sec:risk:r8}
If users are misled due to inadequate information disclosure, it means that $\mathcal{P}_3$ breaks.
Specifically, if a source code verification is passed, the verifying platforms should display the verified information to the public.
Take the Etherscan, the largest Ethereum browser, as an example.
It illustrates the contract name, optimization enabled, and solc version. However, the most important thing is that the contract name does not contain its directory.
Such an inadequate information disclosure may be maliciously utilized.
For example, the verifying requester can put the actual main contract under the path \texttt{./test/test.sol}. When requesting verification, he can attach a seemingly malicious contract with an identical contract name under \texttt{./src/main.sol}, which should not be compiled and verified at all.
On Etherscan pages, because only the contract name is displayed, users may take the one under \texttt{./src/main.sol} as the actual contract for granted. This could severely impact the confidence of the whole project.

\section{Vulnerability Detection \& Exploitation}
\label{sec:detection}

We next present our detection and exploitation methods for each of the vulnerabilities presented.
The consequences of exploiting these vulnerabilities enable malicious users to accomplish competitive verification for benign contracts in ($\mathcal{A}_1$) or source code scams with malicious code ($\mathcal{A}_2$).

\subsection{Vulnerabilities Leading to $\mathcal{A}_1$}
\label{sec:detection:a1}

\subsubsection{Exploitable Compiler Features ($\mathcal{R}_1$)}
\label{sec:detection:a1:r1}
$\mathcal{R}_1$ refers to that the compilation results, i.e., the final generated bytecode, can be controlled (fully or partially) by source code verification requesters, through compiler features like verbatim functions of YUL or loose inline assembly.
Therefore, to detect if $\mathcal{R}_1$ is exploitable in three verifiers, we construct a set of contracts. Each of them is composed of a single fallback function, where it embeds a verbatim function or a piece of loose inline assembly code.
The reason for choosing fallback function is that a fallback function has an empty function signature, thus it is compiled into a piece of bytecode without wrapping other opcodes, like function identifiers~\cite{functionselector}.
Through observing the difference between the opcodes declared in source code and the consequently compiled ones, we can conclude if $\mathcal{R}_1$ is exploitable in verifiers.

\begin{lstlisting}[language=Solidity, caption={The PoC of $\mathcal{R}_1$.}, label={lst:s1}]
contract A_ {
 //target bytecode '608060405260043610610133..'
 function() external payable{
  assembly{      //6080604052
    0x4          //6004
    calldatasize //36
    lt           //10
    tag1         //610133
    ...
  }
 }
}
\end{lstlisting}
\vspace{-0.1in}

\noindent
\textbf{PoC.}
Consider a deployed contract named \texttt{A}, whose runtime code is \texttt{0x608060...}. 
We can compose a contract \texttt{A\_}, as shown in Listing~\ref{lst:s1}, which only contains a fallback function (L4).
To exploit $\mathcal{R}_1$, we firstly translate the victim's, i.e., \texttt{A}'s, runtime code (without metadata code) into a piece of loose inline assembly in the fallback function of \texttt{A\_}, as shown from L4 to L10 at Listing~\ref{lst:s1}.
Then, we compile \texttt{A\_} locally and manually replace the metadata code of \texttt{A\_} to the \texttt{A}'s.
In this way, \texttt{A\_} and \texttt{A} are identical in terms of on-chain runtime code.
Thus, we upload \texttt{A\_} and verify it by providing Listing~\ref{lst:s1}, which should be an exact match certainly\footnote{Note that, the modified metadata code has no influence on the verification result because it is removed before comparing, see \S\ref{sec:source-code-verifier:components:m4}}. Consequently, due to the existence of shortcuts in Etherscan (see Fig.~\ref{fig:module}), \texttt{A} is automatically labeled as verified with the source code of \texttt{A\_}.
As for Sourcify, it allows only verifying the runtime code (see \S\ref{sec:source-code-verifier:components:m4}), leading to an easier attack.
We can directly request verification for \texttt{A} with \texttt{A\_} declared in Listing~\ref{lst:s1}.
After the metadata code is removed, they certainly have an identical runtime code to pass the verification.
Since Blockscout inherits the results of Sourcify, Blockscout is also exploitable.

\noindent
\textbf{Exploitation Conditions.}
Successfully constructing a PoC to exploit $\mathcal{R}_1$ is limited by some conditions.
First, only valid opcodes can be written in loose inline assembly, resulting in the PoC has to be compiled by solc \texttt{$\leq$ 0.4} and has only one piece of metadata.
If solc is greater than \texttt{0.4}, the compiler automatically appends an invalid opcode \texttt{fe} in front of the tailing metadata code. And if there are other metadata code located in the middle of runtime code, it cannot guarantee that such a random hash string can always correspond to valid opcodes.
Second, operands of \texttt{PUSH} cannot be led by zero. In Ethereum, there are 32 types of \texttt{PUSH} which can push 1 to 32 bytes, respectively. However, if we intend to encode opcodes like \texttt{PUSH2 0001} in loose assembly, the compilation automatically optimizes it as \texttt{PUSH1 01}, which changes the deployed runtime code.

\subsubsection{Unchecked Simulated Execution ($\mathcal{R}_2$)}
\label{sec:detection:a1:r2}

As we mentioned in \S\ref{sec:risk:r2}, simulated execution can be regarded as a notorious \texttt{eval()} function, which takes the constructor as input and simulatedly executes opcodes in it.
Thus, to detect $\mathcal{R}_2$, we deliberately insert some functions that can explicitly change control flow in the constructor, like \texttt{call} and \texttt{return}.
To this end, by observing if the control flow of the compiled bytecode changes, we can conclude if $\mathcal{R}_2$ is exploitable.

\begin{lstlisting}[language=Solidity, caption={The PoC of $\mathcal{R}_2$.}, label={lst:s2}]
contract A_ {
 constructor() public {
  // The victim A runtime code
  bytes memory bytecode = hex'608060...';
  assembly {
   return (add(bytecode, 0x20), mload(bytecode))
  }
 }
}
\end{lstlisting}
\vspace{-0.1in}

\noindent
\textbf{PoC.}
Suppose a contract \texttt{A} is deployed on-chain, and we can construct a contract \texttt{A\_} as shown in Listing~\ref{lst:s2}.
At L4, it assigns the runtime code of \texttt{A} to the variable \texttt{bytecode}.
In normal creation code, the runtime code is copied to memory through the opcode \texttt{codecopy}, and a pointer directing to its beginning is returned.
However, at L6, a return is explicitly declared in YUL before the actual \texttt{codecopy} opcode, which is located at the end of the creation code.
In other words, the simulation executes the \texttt{return} at L6 before running into the actual \texttt{codecopy}, leading to an early return of a pointer, which points to the constructed \texttt{bytecode}.
Considering that such an attack can accomplish the \textit{exact match}, we can generate source code files to arbitrary contracts that are not exactly matched yet on Sourcify.

\subsubsection{Incomplete Bytecode Validation ($\mathcal{R}_3$)}
\label{sec:detection:a1:r3}
As we mentioned in \S\ref{sec:risk:r3}, to avoid providing concrete arguments from users, verifiers adopt prefix matching in $\mathcal{M}_4$ when comparing bytecode.
Thus, to detect if $\mathcal{R}_3$ is exploitable in verifiers, we intentionally design a set of contracts, which can generate a prefix of the victim contract. This process can be completed through utilizing loose inline assembly (see \S\ref{sec:detection:a1:r1}) or by composing \textit{abstract contracts}.
Specifically, the abstract contract is a type of contract that can generate an empty bytecode. In other words, an abstract contract can be regarded as a prefix of any contract.
If such contracts can pass the verification of verifiers, $\mathcal{R}_3$ is exploitable.

\begin{lstlisting}[language=Solidity, caption={The PoC of $\mathcal{R}_3$.}, label={lst:s3}]
pragma solidity ^ 0.8.0;
abstract contract LLL{ }
\end{lstlisting}
\vspace{-0.1in}

\noindent
\textbf{PoC.}
To exploit $\mathcal{R}_3$, we choose any victim contract that is unverified.
We claim the code snippet in Listing~\ref{lst:s3} is the corresponding source code, which is an abstract contract.
Because of the problematic implementation of verifiers (e.g., Sourcify~\cite{sourcifystartwith}), such an empty string can be regarded as the prefix of any non-empty contract.
To this end, the source code verification can be bypassed.

\subsection{Vulnerabilities Leading to $\mathcal{A}_2$}
\label{sec:detection:a1}

\subsubsection{Unreliable Client Node ($\mathcal{R}_4$)}
\label{sec:detection:a2:r4}
As we mentioned in \S\ref{sec:risk:r4}, RPCs of client nodes may return unrecognized data, which can be maliciously utilized by attackers. To detect the presence of this vulnerability, a normal method is to deploy multiple client nodes and subtly modify their configuration to break the consensus algorithm to generate unrecognized data, which is out of scope of this work.
Thus, we use an alternative approach to detect the vulnerability of $\mathcal{R}_4$, i.e., through \texttt{create2}. Specifically, if the creation code is unchanged, the deployed address keeps unchanged through \texttt{create2}.
Therefore, to detect if $\mathcal{R}_4$ is exploitable, we firstly compose a contract as a deployer, which has a fixed creation code and takes runtime code as input from arguments. Then, the deployer can deploy runtime code through the PoC in \S\ref{sec:detection:a1:r2}.
Intuitively, if an address can be deployed after the original contract is self-destructed, the verified source code may correspond to the original one, indicating verifiers are exploitable to $\mathcal{R}_4$.

\begin{lstlisting}[language=Solidity, caption={The toolchain of conducting PoC of $\mathcal{R}_4$.}, label={lst:s8}]
contract Deployer {
 bytes public deployBytecode;
 address public deployedAddr;
 function deploy(bytes memory code) public {
  deployBytecode = code;
  address target;
  bytes memory dumperBytecode = hex"{Dumper contract's creation code}";
  assembly {
   target := create2(callvalue, add(0x20, dumperBytecode), mload(dumperBytecode), 0x11)
  }
  deployedAddr = target;}
}
contract Dumper {
 constructor() public {
  Deployer dp = Deployer(msg.sender);
  bytes memory bytecode = dp.deployBytecode();
  assembly {
   return (add(bytecode, 0x20),mload(bytecode))
    }}
}
\end{lstlisting}
\vspace{-0.1in}

\begin{figure}[t]
    \centering
    \includegraphics[width=0.8\linewidth]{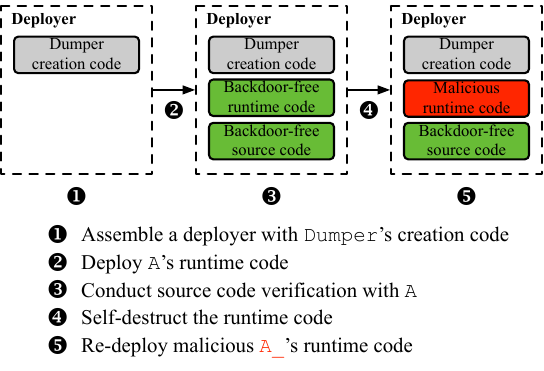}
    \vspace{-0.2in}
    \caption{Overview of the PoC of $\mathcal{R}_4$.}
    \vspace{-0.2in}
    \label{fig:s8-overview}
\end{figure}

\noindent
\textbf{PoC.}
Fig.~\ref{fig:s8-overview} and Listing~\ref{lst:s8} illustrate how a fraud can be conducted through exploiting $\mathcal{R}_4$.
In the preparation stage, i.e., \textbf{step 1}, we first compile the \texttt{Dumper} declared at L14 in Listing~\ref{lst:s8}, extract its creation code and assign it to \texttt{dumperBytecode} at L7. Thus, taking advantage of Solidity, the contract \texttt{Deployer} can be seen as a general deployer. This means that any runtime code passed through \texttt{code} at L4 can be deployed at an unchanged address with such a creation code.
During exploiting $\mathcal{R}_4$ at \textbf{step 2}, we first compile a backdoor-free contract \texttt{A}, and call the function \texttt{deploy} of \texttt{Deployer} at L4 with \texttt{code} as \texttt{A}'s runtime code. The \texttt{create2} at L9 deploys this contract on an address.
Then, at \textbf{step 3}, we provide the corresponding source code files of \texttt{A} and ask for a verification. After the verification completes, we call the function \texttt{selfdesctruct} in \texttt{A} to make this address available, as shown in \textbf{step 4}.
To complete the exploitation, we again compose another evil contract \texttt{A\_}, and pass its runtime code in \texttt{deployer}.
Due to the characteristic of \texttt{create2}, such an evil contract is also deployed on the identical address to \texttt{A}.
Consequently, the verification results of \texttt{A} are mistakenly bound on the actual executed contract, i.e., \texttt{A\_}.

\subsubsection{Unverified Linked Libraries ($\mathcal{R}_5$)}
\label{sec:detection:a2:r5}
As we mentioned in \S\ref{sec:risk:r5}, the implementation of invoked linked libraries is not recursively verified, which breaks $\mathcal{P}_2$.
To detect if $\mathcal{R}_5$ is exploitable, we firstly deploy a contract that invokes a linked library where it poses malicious behaviors. Then, we verify the contract by uploading a library with identical function signature but different implementation.
If the verification can pass, it means $\mathcal{P}_2$ is violated and malicious users can conduct fraud through exploiting $\mathcal{R}_5$.

\begin{lstlisting}[language=Solidity, caption={The PoC of $\mathcal{R}_5$.}, label={lst:s4}]
pragma solidity ^0.8.0;
contract A{ // caller
  uint totalsupply = 0;
  function is_zero() public view returns(bool){
    L.check(totalsupply);
    // compiled to: L.delegatte(abi("check(uint)", totalsupply))
    return true;
  }
}
library L{ // linked library
  function check(uint balance) public view{
    require(balance == 0);
  }
}
\end{lstlisting}
\vspace{-0.1in}

\noindent
\textbf{PoC.}
Listing~\ref{lst:s4} illustrates a call to a linked library. We can see from L5 that \texttt{L} is a linked library, and \texttt{check} is the callee function. This is equivalent to the statement at L6, i.e., only the address of \texttt{L} and the signature of \texttt{check} is considered by the caller contract.
To this end, we can conduct fraud by utilizing this feature.
Specifically, we first update Listing~\ref{lst:s4} by replacing L12 to \texttt{selfdestruct(msg.caller)} that can destroy the innocent caller's contract.
Then, we compile and deploy this malicious contract on-chain.
To conduct exploitation, we provide the seemingly normal Listing~\ref{lst:s4} during the source code verification, which should pass the verification because the contract \texttt{A} is unchanged, and the signature of the callee function \texttt{check} in the linked library also keeps intact.
Consequently, the malicious code is covered by a fake but valid source code verification.

\subsubsection{Mislabeled Bytecode ($\mathcal{R}_6$)}
\label{sec:detection:a2:r6}
Considering user-friendliness and to avoid unnecessary verification failure, some bytecode is labeled and extracted before comparison in $\mathcal{M}_4$.
To detect if $\mathcal{R}_6$ is exploitable, two types of bytecode should be considered, \textit{metadata} and \textit{linked library placeholders}.

As for labeling metadata code, two methods are raised before, i.e., \textit{regex matching} and \textit{differential extracting}.
In 2021, a white hat utilized the buggy regex matching to intentionally mark a piece of runtime code as metadata code, where he hid a backdoor~\cite{samczsunvul}. It proves the unreliability of this strategy.
Thus, Blockscout has proposed the idea of differential extracting. 
According to its implementation~\cite{bdiffextra}, the differential extracting process can be summarized as the following steps.
\begin{itemize}
\item It compiles the given source code files according to configuration items, which is denoted as $c$.
\item Blockscout intentionally adds a pre-defined and useless configuration item to make the compiled bytecode changed as $c'$. The item is a file named \texttt{SOME\_TEXT\_USED\_AS\_FILENAME}, which contains a library whose name is calculated dynamically according to the name of contracts in the main file~\cite{bdiffimpl}.
\item By comparing $c$ and $c'$, Blockscout can identify an index, denoted as $i$, where the first difference occurs. Based on $i$, it traverses backward and forward to find a byte string that follows the pattern of a valid metadata code (see \S\ref{sec:risk:r6}).
\item The above indexing and metadata identifying process repeat till all metadata code fields are considered.
\end{itemize}
Therefore, to detect if this type of $\mathcal{R}_6$ is exploitable, we compose a contract, in which we deliberately invoke the library imported by Blockscout. Then, between $\mathcal{M}_3$ and $\mathcal{M}_4$, we compare the compilation results of the Blockscout generated one and the normal one. If some runtime logic is missed, it means that runtime logic is mislabeled, violating $\mathcal{P}_2$.

As for labeling linked library placeholders, we examine if this type of $\mathcal{R}_6$ is exploitable on Sourcify because it is the only that does not require the addresses of invoked linked libraries.
Specifically, the process can be abstracted as:
\begin{itemize}
\item The verifier identifies the first appeared placeholder, records its offset, and extracts it as a \textit{regex pattern}.
\item The verifier locates the bytes from on-chain bytecode according to the recorded offset.
\item Among all placeholders, it matches the ones according to the regex pattern, and replaces them with the identified address.
\item The above processes repeat till no placeholders exist.
\end{itemize}
This implementation is safe when solc is greater than \texttt{0.4}, where the placeholder is a 34 bytes hash with fixed prefix and suffix.
However, the situation turns opposite when solc is \texttt{0.4}, where the placeholder is a string like:
\begin{align*}
\texttt{\_\ \_\ }\ filePath\ \texttt{:}\ libName\ \texttt{\_\ \_}
\end{align*}
, where \textit{filePath} and \textit{libName} are strings explicitly declared in the source code.
Because solc does not require the format of \textit{filePath}, e.g., ended by \texttt{.sol}, attackers can construct a value to utilize Sourcify's regex matching mechanism.
Therefore, to detect if $\mathcal{R}_6$ is exploitable and feasible on Sourcify, we construct a contract whose \textit{filePath} and \textit{libName} are deliberately set as strings following regex grammar. Then, after $\mathcal{M}_3$ of Sourcify, we observe if the forged \textit{filePath} and \textit{libName} can hit any valid runtime logic and replace it with on-chain bytecode. If it is, it means attackers can construct malicious contracts to arbitrarily label runtime logic to intentionally remove it before comparison in $\mathcal{M}_4$.

The following \textbf{PoC \#1} and \textbf{PoC \#2} illustrate how to mislabel metadata and linked library placeholders to hide malicious code, respectively.

\begin{figure}[t]
    \centering
    \includegraphics[width=1\linewidth]{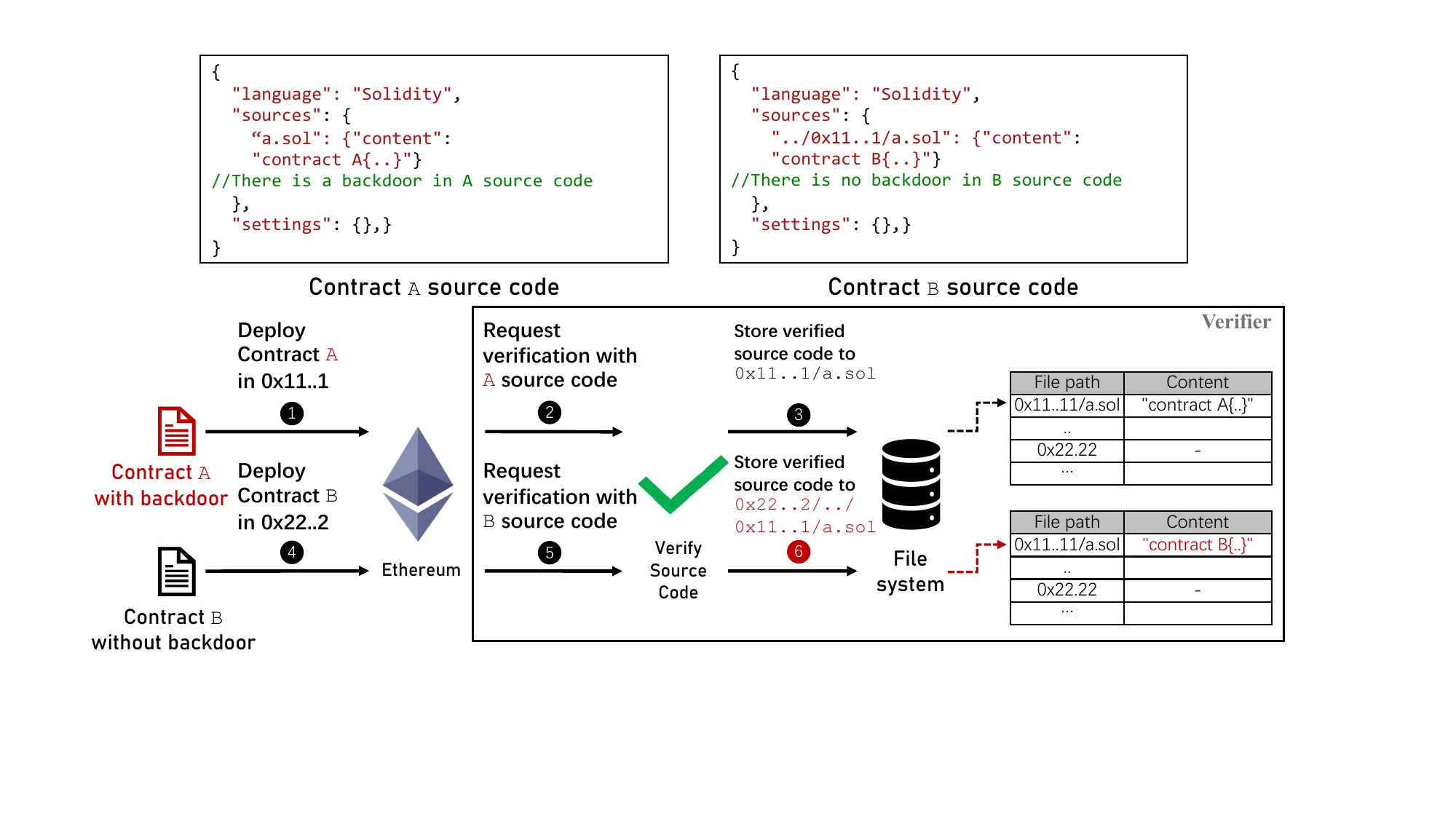}
    \vspace{-0.3in}
    \caption{Overview of the PoC \#1 of $\mathcal{R}_6$.}
    \vspace{-0.2in}
    \label{fig:s6-overview}
\end{figure}

\noindent
\textbf{PoC \#1.}
Fig.~\ref{fig:s6-overview} illustrates a concrete example of how to exploit this type of $\mathcal{R}_6$.
Specifically, we compose a piece of source code, as shown in the top of Fig.~\ref{fig:s6-overview}. 
We deliberately import a file named \texttt{SOME\_TEXT\_USED\_AS\_FILENAME}, and assign the library (\texttt{L\_}), which is defined in the imported file, to a variable.
At \textbf{step 1 \& 2}, we compile the source code, and change an opcode near the library placeholder to \texttt{FF}, i.e., \texttt{selfdestruct}. Then, we deploy such a modified bytecode on-chain.
After that, we request a source code verification through Blockscout.
Because the library \texttt{L\_} is defined in the imported strange file, which is inserted once Blockscout updates configuration items during conducting differential extracting, the bytecode at that location changes.
According to the implementation adopted by Blockscout we mentioned before, it should mark the \texttt{L\_} nearby area, including the malicious \texttt{FF}, as metadata code and remove it, which is normal runtime logic actually.
Consequently, because other parts of bytecode are still identical to the on-chain ones, it can pass the verification, while users are unaware of there is a dangerous \texttt{selfdestruct} hidden in the deployed contract.

\begin{lstlisting}[language=Solidity, caption={The PoC \#2 of $\mathcal{R}_6$.}, label={lst:case5}]
pragma solidity ^0.4.0;
import "./$.{37}|2{40}|"; // file path
contract A {
 address constant public owner = address(0x222..22);
 uint public b;
 function c() public{
   b = foo.bar();
}}

library foo{ // lib name
 function bar() public pure returns(uint) {
  return 1;
}}
\end{lstlisting}
\vspace{-0.1in}

\noindent
\textbf{PoC \#2.}
Listing~\ref{lst:case5} illustrates a malicious smart contract.
The L2 and L10 correspond to \textit{filePath} and \textit{libName}, respectively. Thus, a placeholder is generated as\footnote{The $\ldots$ refers to underline characters to make the whole placeholder as 40 bytes long.}:
\begin{align*}
&\texttt{\_\_\$.\{37\}|2\{40\}|:foo\_\ldots\_\_}
\end{align*}
To this end, Sourcify takes the above placeholder as a regex pattern. Specifically, the \texttt{\_\_\$.\{37\}} can match the current placeholder itself, but the \texttt{2\{40\}} can match the address declared at L4.
Suppose this placeholder locates at the offset $o$.
We can place any 40-byte sequences at $o$ on the on-chain contract, like an address or even a backdoor. Therefore, through the constructed regex pattern, Sourcify mistakenly replaces the address at L4 to a predefined byte sequence.

\subsubsection{Path Traversal Risk ($\mathcal{R}_7$)}
\label{sec:detection:a2:r6}
In \S\ref{sec:risk:r7}, we mentioned that users can input arbitrary file paths through the input JSON when requesting source code verification services, which may result in path traversal or file overwriting. To detect if $\mathcal{R}_7$ is exploitable, we construct several JSON files, within which the path of source code files are declared in a malicious path traversal way, like \texttt{../} to refer to its parent directory.
Then, we try to access the directory to see if the source code files are uploaded to that invalid directory.

\noindent
\textbf{PoC.}
Constructing this PoC is intuitive. Suppose there is a contract with backdoors named \texttt{A}. After the deployment and source code verification, let us assume its source code is stored in \texttt{0x12..fe/source/A.sol}.
To fool users, we can deploy another contract without the backdoors, named \texttt{A\_} for instance.
However, during performing the source code verification, we claim the path of \texttt{A\_} is:
\begin{align*}
\texttt{../../0x12..fe/source/A.sol}
\end{align*}
To this end, the verifier stores the source code of \texttt{A\_} at the place where it should be used to store \texttt{A}.
When users visit verifiers to examine the source code files of \texttt{A}, they are fooled by the implementation of \texttt{A\_}.

\subsubsection{Inadequate Information Disclosure ($\mathcal{R}_8$)}
\label{sec:detection:a2:r8}

Inadequate information disclosure can be abused to mislead users. For example, in Blockscout and Etherscan, the field \textit{compile target} only consists of \textit{contract name}. Considering solc allows compilation with multiple source code files, lack of \textit{source file paths} cannot eliminate the ambiguity when referring a contract from the ones with identical name.
The detection of $\mathcal{R}_8$ is intuitive, i.e., we can upload multiple contracts with an identical name to verifiers. Then, we observe if users can distinguish which one is the main contract. If they can not be distinguished through displayed information, we conclude that this verifier is exploitable to $\mathcal{R}_8$.

\noindent
\textbf{PoC.}
We have deployed two contracts, named \texttt{test.sol} and \texttt{main.sol}, respectively. Both of them follow a simplified ERC-721 standard~\cite{erc721} to issue tokens. The only difference between \texttt{test.sol} and \texttt{main.sol} is the \texttt{totalSupply}, i.e., the field refers to the maximum amount of tokens that is allowed to be minted. The value of \texttt{totalSupply} in \texttt{test.sol} is much higher than the one of \texttt{main.sol}.
We deploy \texttt{test.sol} firstly, and verify it by uploading both files simultaneously. 
To fool users further, we even put the \texttt{test.sol} under a deeper and a more confusing directory. 
Consequently, users may be fooled by a seemingly limited edition of the one shown in \texttt{main.sol}.

\section{Evaluation}
\label{sec:evaluation}

\subsection{Experimental Setup \& Ethical Considerations}
\label{sec:evaluation:test-env}

In this work, we target the most popular Ethereum smart contract verification services, including Etherscan, Sourcify, and Blockscout.
For all these three verifiers, they provide the verification services on both Ethereum mainnet and testnet. For ethical considerations, all evaluations are conducted on the verifiers that are responsible for the \textit{goerli} testnet, which is independent of the Ethereum mainnet.

We have carefully designed methods to detect the exploitable vulnerabilities.
Specifically, for the $\mathcal{A}_1$-related vulnerabilities, i.e., $\mathcal{R}_1$ to $\mathcal{R}_3$, we randomly sample 10 widely-used contracts with source code from mainnet as victim contracts, and deploy them on the testnet. After that, according to the victim's bytecode, we construct PoCs to try to pass the verification. Note that a successful verification does not imply the $\mathcal{A}_1$ consequence, as we need to ensure that the forged source code cannot be replaced.
Therefore, we further verify these victim contracts with their original source code by requesting the verification services.
If the re-verification fails, we conclude that the testing verifier is exploitable.
For the $\mathcal{A}_2$-related vulnerabilities, we firstly construct some contracts with backdoors, like conducting \texttt{selfdestruct} or transferring to other users, compile and deploy them on the testnet. Then, according to PoCs raised from $\mathcal{R}_4$ to $\mathcal{R}_8$, we try to construct source code without such unexpected behaviors, and bypass the source code verification.
If it succeeds and no obvious warnings raised by verifiers, we can conclude that the targeting verifier is exploitable.

Finally, after confirming the exploitability of vulnerabilities, we conduct timely disclosure to impacted verifiers within 30 minutes. On the one hand, it benefits a timely fixup on exploitable vulnerabilities. On the other hand, it gives little chance for malicious users to replay this exploitation by observing our uploaded source code files.

\subsection{Detecting Results}
\label{sec:evalulation:results}

\begin{table}[]
\centering
\caption{Overall vulnerability detection results. For each vulnerability, -- and \xmark\ refer to infeasible (safe) and exploitable, respectively. * refers to the official teams have confirmed our reported vulnerabilities, and \cmark\ indicates it has been patched after our timely disclosure.}
\label{table:relation}
\begin{tabular}{@{}ccccc@{}}
\toprule
\textbf{Consequence}             & \textbf{\begin{tabular}[c]{@{}c@{}}Exploitable\\ Vulnerailities\end{tabular}} & \textbf{Etherscan} & \textbf{Sourcify} & \textbf{Blockscout} \\ \midrule
\multirow{3}{*}{$\mathcal{A}_1$} & $\mathcal{R}_1$                                                               & \xmark             & \xmark *            & \xmark *              \\
                                 & $\mathcal{R}_2$                                                               & --                  & \xmark *\ (\cmark)            & \xmark *\ (\cmark)              \\
                                 & $\mathcal{R}_3$                                                               & --                  & \xmark *\ (\cmark)            & \xmark *\ (\cmark)              \\ \midrule
\multirow{5}{*}{$\mathcal{A}_2$} & $\mathcal{R}_4$                                                               & \xmark             & \xmark *            & \xmark *              \\
                                 & $\mathcal{R}_5$                                                               & \xmark             & \xmark *            & \xmark *              \\
                                 & $\mathcal{R}_6$                                                               & --                  & \xmark *\ (\cmark)                 & \xmark *\ (\cmark)              \\
                                 & $\mathcal{R}_7$                                                               & --                  & \xmark *\ (\cmark)            & \xmark *\ (\cmark)              \\
                                 & $\mathcal{R}_8$                                                               & \xmark             & --                 & \xmark *              \\ \bottomrule
\end{tabular}
\vspace{-0.1in}
\end{table}

\noindent \textbf{Overall Result.}
Table~\ref{table:relation} presents the overall results. Surprisingly, all these popular verifiers are vulnerable, which can be abused by attackers. In total, we have uncovered 19 vulnerabilities that can be exploited. Etherscan is vulnerable to 4 kinds of vulnerabilities, while Sourcify poses the risks of seven types of vulnerabilities, and Blockscout can be exploited by all types of vulnerabilities.

\noindent \textbf{Further Exploration.}
As Sourcify and Blockscout are vulnerable to most kinds of vulnerabilities, we next deep dive into them. For Sourcify, its designs for user-friendliness import challenges to the verifier. For example, Sourcify does not require users to provide the addresses of invoked linked libraries, it tries to replace placeholders according to a regex pattern, which is also embedded in the source code. As for Blockscout, it is vulnerable to all kinds of attacks. Recall the shortcut we mentioned in \S\ref{sec:source-code-verifier:shortcuts}, which can be seen as an amplifier to expand the exploitable scope of PoCs. For example, if a vulnerability is exploitable in Sourcify, we also consider it to be exploited in Blockscout because the latter one recognizes the results generated by the former one.
Through the shortcut, Blockscout inherits the malicious verification results of exploiting $\mathcal{R}_1$, $\mathcal{R}_2$, $\mathcal{R}_3$, $\mathcal{R}_4$ and $\mathcal{R}_7$ from Sourcify.
Besides the above, importantly, Blockscout and Sourcify's transparency enables more insightful security analysis compared to Etherscan.

\noindent \textbf{Vulnerability Patching.}
We disclose the vulnerabilities to verifiers timely. Among the 19 uncovered vulnerabilities, 15 of them have been confirmed by the official teams, and 8 vulnerabilities have been fixed. For Sourcify and Blockscout, we observed that the vulnerabilities of $\mathcal{R}_2$, $\mathcal{R}_3$, $\mathcal{R}_6$, and $\mathcal{R}_7$ were fixed within 12 hours.
However, as suggested in Table~\ref{table:relation}, we find that four kinds of vulnerabilities are almost exploitable in all verifiers by the time of this writing. 
By contacting the official teams of the verifiers, we summarize the following challenges. Specifically, as for $\mathcal{R}_1$, loose inline assembly is a necessary feature that is widely adopted by normal contracts to improve its runtime efficiency~\cite{inlinepractice}. Directly disabling it shall affect the usability of verifiers.
As for the other three kinds of vulnerabilities ($\mathcal{R}_4$, $\mathcal{R}_5$, and $\mathcal{R}_8$), the verification services thought that the users should raise awareness of the scams, e.g., they should pay special attentions to whether a smart contract invokes a malicious linked library or has a suspicious accompanying file with identical name, etc. Thus, they are considering adding new features to raise warnings for users in recent updates.

\begin{table}[]
\centering
\caption{Affected Contracts Statistical Table.}
\label{table:1}
\begin{tabular}{@{}ccc@{}}
\toprule
\textbf{Consequence}                      & \textbf{\begin{tabular}[c]{@{}c@{}}Exploitable\\ Vulnerailities\end{tabular}}         & \textbf{\# Afftected Contracts}                                                                                     \\ \midrule
\multirow{6}{*}{$\mathcal{A}_1$} & $\mathcal{R}_1$ & 49,598                                                                                          \\ \cmidrule(l){2-3} 
                                 & $\mathcal{R}_2$ & \begin{tabular}[c]{@{}c@{}}partial verified / unverified\\ contracts ($\sim$58.9M)\end{tabular} \\ \cmidrule(l){2-3} 
                                 & $\mathcal{R}_3$ & \begin{tabular}[c]{@{}c@{}}unverified contracts\\ ($\sim$58.9M)\end{tabular}                    \\ \midrule
\multirow{3}{*}{$\mathcal{A}_2$} & $\mathcal{R}_4$ & 2                                                                                              \\ \cmidrule(l){2-3} 
                                 & $\mathcal{R}_5$ & 244                                                                                             \\ \bottomrule
\end{tabular}
\vspace{-0.1in}
\end{table}

\subsection{Contracts Affected by $\mathcal{A}_1$-related Vulnerabilities}
\label{sec:evaluation:a1}

\subsubsection{Method}
Among all three $\mathcal{A}_1$-related vulnerabilities, exploiting $\mathcal{R}_2$ and $\mathcal{R}_3$ have no prerequisites. Thus the number of influenced smart contracts are all smart contracts that are not exactly matched yet and unverified, respectively.
As for $\mathcal{R}_1$, due to the restriction we mentioned in \S\ref{sec:detection:a1:r1}, we compose a simple and effective SQL query statement to filter out all possible victim contracts, as shown in Listing~\ref{lst:i1}.
It consists of three sets of \texttt{like} operators, based on the patterns we summarized.
Specifically, first, to narrow down the scope to all contracts compiled from solc \texttt{= 0.4}, we heuristically adopt the pattern at L3, which is a magic number located in metadata code that is used by \texttt{0.4} solc.
L5 and L6 restrict the operand of each \texttt{PUSH} operators has no leading zeros.
Note that, there are 32 \texttt{PUSH} operators defined by Solidity, we only illustrate \texttt{PUSH2} and \texttt{PUSH3} in Listing~\ref{lst:i1}.
Of course, because this attack is only applicable on contracts with at most a single piece of metadata code, it is filtered by L9.
To be specific, L9 consists of 4 bytes, where \texttt{0029} and \texttt{a165} are standard ending and starting of metadata, respectively. If such a byte string does not exist, it can be greatly guaranteed that there do not exist no more than one metadata.

\begin{lstlisting}[language=sql, caption={SQL query for counting contracts affected by exploiting $\mathcal{R}_1$.}, label={lst:i1}]
SELECT * FROM `crypto_ethereum.contracts` 
    WHERE 
    bytecode like '%a165627a7a7230%' and
    -- 0.4 version
    bytecode not like '%6100%' and -- PUSH2 00XX
    bytecode not like '%620000%' and -- PUSH3 0000XX
    ...
    -- filter out operands with leading zero in PUSHX
    bytecode not like '%0029a165%'
    -- multi-metadata
\end{lstlisting}
\vspace{-0.1in}

\subsubsection{Result}
As shown in Table~\ref{table:1}, We finally identify more than 49k contracts that are influenced by exploiting $\mathcal{R}_1$. Among them, by identifying the function signatures, 25K and 2,440 contracts are suspected to follow ERC20 and ERC721, respectively, which may lead to huge financial and ecological impact if they are verified by an irrelevant piece of source code.
In addition, all unverified solidity contracts in Sourcify (i.e., around 58.9M) are affected by $\mathcal{R}_2$ and $\mathcal{R}_3$, where the former one even includes the partial verified ones (around to 9K contracts). Moreover, as we clarified in \S\ref{sec:intro}, several types of downstream applications rely on the source code verification services.
If millions of contracts can be competitively verified by forged smart contracts, the effectiveness of these downstream applications are severely impacted.
We illustrate a case study in our artifact~\cite{allinone}, where Slither~\cite{feist2019slither}, a well-known vulnerability detector on Ethereum smart contracts with source code, is utilized by forged source code files, resulting in arbitrary file overwriting in its hosting environment.

\subsection{Contracts Affected by $\mathcal{A}_2$-related Vulnerabilities}
\label{sec:evaluation:a2}
As shown in Table~\ref{table:relation}, five vulnerabilities are exploitable to result in source scam against users.
Therefore, we try to measure if there exists any contracts that have already performed such tricks on users.

\subsubsection{Method}
To the best of our knowledge, among all these vulnerabilities that are related to the consequence $\mathcal{A}_2$, $\mathcal{R}_6$ to $\mathcal{R}_8$ are zero-day vulnerabilities uncovered by us. No existing corresponding instances are found on Ethereum mainnet.
Thus, we mainly focus on the existence of instances that exploit $\mathcal{R}_4$ and $\mathcal{R}_5$.
Based on tintinweb~\cite{ortnersmart}, a dataset consisting of all verified contracts on Ethereum, we performed a series of rule-based detection to filter contracts that may have exploited these two vulnerabilities to fool users.
Specifically, as for identifying $\mathcal{R}_4$, for each verified contract, we iterate its transactions to identify if there are more than one successful contract deployment.
To identify $\mathcal{R}_5$, we first filter out all verified contracts that have a call relationship to linked libraries. Then, we perform a difference checking between the uploaded linked libraries under the caller address and the linked libraries declared by the address if they are also verified.
Thus, we can identify if there is an attacker who uploads a forged linked library that has an identical signature of the callee function.

\subsubsection{Result}
As shown in Table~\ref{table:1}, we finally find 2 and 244 contracts whose developers are suspected to utilize $\mathcal{R}_4$ and $\mathcal{R}_5$ for conducting a scam. For the 2 cases under $\mathcal{R}_4$, after a manual recheck, we confirm that one case has performed a fake proposal attack against Tornado.cash~\cite{tornadocash}.
Specifically, by exploiting $\mathcal{R}_4$, its attacker associates a piece of source code that has gained the user's trust to a malicious proposal.
The proposal ultimately resulted in a \$750K financial loss for Tornado.cash~\cite{tornadoattack}.
As for the 244 cases that are related to $\mathcal{R}_5$, we confirm that at least 6 of them invoked malicious libraries but uploaded with benign ones. For example, an attack targeted the Saddle Finance~\cite{saddlefinance}, a well-known decentralized exchange. Interestingly, its actually invoked linked library is a deprecated version, which has a vulnerability in the price calculation that can be utilized by malicious users to exchange a small number of token for another bunch of token at a rate deviated from the market. However, the developer of Saddle Finance has uploaded the updated and bug-free version of the invoked library, which covers the vulnerability hidden in the library.
Eventually, this vulnerability has been exploited and led to a financial loss of \$7.2M.

\section{Related Work}

\noindent
\textbf{Blockchain Security.}
Currently, a great deal of research work is focusing on the security of the whole blockchain ecosystem~\cite{peng2018all,chen2017adaptive,li2021strong,heilman2015eclipse,he2021eosafe,wust2016Ethereum,yang2021finding,biryukov2019security,chen2022tyr,chen2022wasai}. While lots of them are delving into the security risks of various modules in the blockchain ecosystem, there is still a void when targeting source code verification services.

\noindent
\textbf{Ethereum Scam.}
In recent years, with the increasing financial value embedded in the Ethereum contracts, the research on scam in the Ethereum ecosystem has become a hot spot~\cite{xia2022challenges,xia2020characterizing,chen2021sadponzi,torres2019art,cheng2019towards,gao2020tracking,xia2021trade}. For example, Chen et al.~\cite{chen2021sadponzi} propose a heuristic-guided symbolic execution technique to identify Ponzi scheme contracts in Ethereum. However, there is little discussion about if scams can be performed by a mismatched source code and deployed bytecode.

\noindent
\textbf{Ethereum Smart Contract Analysis.}
In terms of Solidity contract security, there is a lot of work~\cite{grieco2020echidna,durieux2020empirical,mossberg2019manticore,feist2019slither,ferreira2020smartbugs,tikhomirov2018smartcheck,chen2020soda,akca2019solanalyser,hajdu2020solc,lin2022solsee} focus on contract vulnerability detection based on source code. For instance, Feist et al.~\cite{feist2019slither} build a static analysis framework for solidity source code, called Slither, to provide support for bug detection in Ethereum smart contracts. However, all of them assume the fetched source code files are correct and not consider the risks brought by source code verifiers.

\section{Discussion}

\noindent \textbf{Threats to Validity.} First, there might be more kinds of vulnerabilities related to smart contract verification services, which are not covered in this work. Etherscan is close-sourced and we can only conduct a black-box analysis against it. Nevertheless, we argue that we have comprehensively performed a code audit on Blockscout and Sourcify against the intuitive security properties, and summarized the vulnerabilities related to almost all the components of verifiers. We believe this work can remind developers and security researchers of this previously under-studied direction. Second, beyond the detection principles for vulnerabilities, we also illustrate the corresponding PoCs for them.
However, we need to mention that these exploitations are not the only way to tamper verifiers.
For example, in $\mathcal{R}_2$ (see \S\ref{sec:detection:a1:r2}), the attacker could adopt \texttt{return} in YUL inline assembly to exploit the simulated execution mechanism. Furthermore, he could also utilize the jump statement in loose inline assembly~\cite{jop}.
Moreover, several vulnerabilities can be combined by attackers to perform a more covert attack.
In a nutshell, we have presented these vulnerabilities in the most understandable way, but developers should never take the risks hidden in verifiers lightly. In this work, we have considered the first three mainstream verifiers in Ethereum, and identified several exploitable vulnerabilities.
There do exist other verifiers in Ethereum, e.g., Tenderly~\cite{tenderly}. We argue that our detecting principles can be directly applied to them.

\noindent \textbf{Mitigation.}
As shown in Table~\ref{table:relation}, eight vulnerabilities have been fixed timely, which are relatively straightforward.
For example, in $\mathcal{R}_2$, attackers can construct arbitrary bytecode that is returned by simulated execution. To fix this issue, Sourcify directly removes this feature. It fetches the on-chain bytecode at the same offsets and assigns their values to the corresponding immutable variables.
Though the immutable values are not verified at all, it is fairly a trade-off when considering the devastating results of $\mathcal{A}_1$.
The fixes for other vulnerabilities generally follow this principle, i.e., disabling the corresponding features to eliminate the nondeterminism brought by some automatic processes.
For instance, against $\mathcal{R}_3$, Sourcify starts to examine if the remaining part can be decoded as valid arguments.

\section{Conclusion}
We explored the security issues of smart contract verification services in this work, an uncharted direction of the Ethereum ecosystem. We have depicted the design and implementation of smart contract verification services, summarized their security properties, observed eight kinds of vulnerabilities, and proposed effective detection and attack methods. Our exploration has uncovered 19 exploitable vulnerabilities in popular smart contract verification services, posing a great impact to millions of smart contracts in the ecosystem. Our results encourage our research community to invest more efforts into the under-studied directions.

\balance
\bibliographystyle{IEEEtranS}
\bibliography{sample-base}
%



\end{document}